\documentclass{PoS} 
 
\usepackage{stmaryrd}

\def\IC{\mathbb{C}}

\def\IZ{\mathbb{Z}}
\def\IR{\mathbb{R}}
\def\ID{\mathbb{D}}

\def\IJ{\mathbb{J}}

\def\IX{\mathbb{X}}

\def\IY{\mathbb{Y}}

\def\Ilb{\llbracket}
\def\Irb{\rrbracket}

\newcommand{\pp}{{=\!\!\!\!\!|}}

\newcommand\cmp[3]{{\it Commun.\ Math.\ Phys.\ }{\bf #1} (#2) #3}
\newcommand\jhep[3]{{\it J. High Energy Phys.\ }{\bf #1} (#2) #3}

\newcommand\npb[3]{{\it Nucl.\ Phys.\ }{\bf B #1} (#2) #3}
\newcommand\plb[3]{{\it Phys.\ Lett.\ }{\bf B #1} (#2) #3}

\newcommand\forp[3]{{\it Fortschr.\ Phys.\ }{\bf #1} (#2) #3}
\newcommand\hepth[1]{{\tt hep-th/#1}}

\title{ $N = (2, 2)$ Non-Linear $\sigma$-Models: A Synopsis} 
 
\ShortTitle{N = (2, 2) Non-Linear $\sigma$-Models: A Synopsis} 
 
\author{\speaker{Alexander Sevrin}\thanks{Also at the Physics Department, Universiteit Antwerpen,
Campus Groenenborger, 2020 Antwerpen, Belgium.} $\ $and Daniel C. Thompson 
         \\ 
Theoretische Natuurkunde, Vrije Universiteit Brussel\\ and\\The International Solvay Institutes\\
Pleinlaan 2, B-1050 Brussels, Belgium\\
        E-mail: \email{Alexandre.Sevrin@vub.ac.be} and \email{Daniel.Thompson@vub.ac.be}} 
 
\abstract{We review $N=(2,2)$ supersymmetric non-linear $\sigma$-models in two dimensions and their
relation to generalized K\"ahler and Calabi-Yau geometry. We illustrate this with an explicit non-trivial example.} 
 
\FullConference{ Proceedings of the Corfu Summer Institute 2012 \\ 
		 September 8-27, 2012\\ 
		 Corfu, Greece} 
 
\begin{document} 
 \section{Introduction}
Moduli stabilization, the AdS-CFT correspondence and string phenomenology all require the study of strings in non-trivial backgrounds. The method of choice to do so uses the supergravity approximation. In order to go beyond supergravity one is currently forced to use $d=2$ conformal field theories, Landau-Ginzburg type descriptions, the pure spinor formalism or the $\sigma$-model worldsheet description. The latter is limited to non-trivial NSNS backgrounds only. Particularly interesting are non-linear $\sigma$-models in two dimensions with $N=(2,2)$ supersymmetry \cite{Gates:1984nk},  \cite{Howe:1985pm} which play a central role in the worldsheet description of
type II string theories in non-trivial NSNS backgrounds. Even at the classical level these models exhibit a rich geometric structure which is the main topic of the current review paper. 

This paper  closely follows the line of the talk. In the next section we introduce non-linear $\sigma $-models and identify the geometries which allow for an $N=(2,2)$ supersymmetry. Subsequently we introduce Hitchin's generalized K\"ahler geometry, the proper setting for the above mentioned models. After this we present the full off-shell description of these models in $N=(2,2)$ superspace. Having an $N=(2,2)$ superspace description allows one to perform various (T-)duality transformations keeping the supersymmetries manifest. These transformations typically alter the nature of the superfields and we provide an exhaustive list of them. Finally we illustrate all of this using a simple example: the Wess-Zumino-Witten model on $SU(2)\times U(1)$. We end with conclusions and summarize our conventions in the appendix.

\section{From $N=(1,1)$ to $N=(2,2)$ supersymmetry}
In the absence of boundaries, a non-linear $\sigma$-model in two dimensions is classically fully characterized by its $D$-dimensional target manifold $ {\cal M}$,
a metric on the target manifold $g$ and a closed three-form $H$, $dH=0$. The latter is known as the torsion or the NS-NS
form and can be locally written as  $H=d\,b$ with $b$  the Kalb-Ramond two-form. The action is then given by\footnote{Our conventions are given in appendix A.},
\begin{eqnarray}
{\cal S}=\int d^2 \sigma\, \partial _\pp\,\, x^a\big(g_{ab}+b_{ab}\big) \partial_= x^b\,,\label{ac0}
\end{eqnarray}
where $x^a$, $a\in\{1,\cdots,D\}$, is a set of local coordinates on ${\cal M}$. 

Without adding any further geometric structure  this action may be supersymmetrized provided that the number of supersymmetries $N$ is bounded by $ N\leq (1,1)$. In this paper we will exclusively focus on $N=(1,1)$ and $N=(2,2)$. The $N=(1,1)$ supersymmetric version of eq.~(\ref{ac0}) in $N=(1,1)$ superspace is simply,
\begin{eqnarray}
{\cal S}=4\int d^2 \sigma \, d^2 \theta \,D_+x^a\left(g_{ab}+b_{ab}\right) D_-x^b\,.\label{an11}
\end{eqnarray}
Integrating over $\theta^+$ and $\theta ^-$ we get the action in ordinary space,
\begin{eqnarray}
&&{\cal S}=\int d^2 \sigma\,\big(\partial_\pp \,\,x^a(g_{ab}+b_{ab})  \partial_= x^b+
 2i\, \psi ^a_+\,g_{ab}\,\nabla_=^{(+)} \psi ^b_+
+2i\, \psi ^a_-\,g_{ab}\,\nabla_\pp^{(-)} \psi ^b_- \nonumber\\
&&\qquad +\psi _-^a \psi _-^b \,R^{(-)}_{abcd} \,\psi _+^c \psi _+^d
+ 2 (F^a-i\, \Gamma ^{\ a}_{(-)cd} \psi ^c_- \psi ^d_+ )g_{ab}
(F^b-i \,\Gamma ^{\ b}_{(-)ef} \psi ^e_- \psi ^f_+ )\big)\,,\label{acnorsp}
\end{eqnarray}
where we introduced covariant derivatives $\nabla^{(+)}$ and $\nabla^{(-)}$ which use the (Bismut) connections $\Gamma_{(+)}$ and $\Gamma_{(-)}$ resp.,
\begin{eqnarray}
\Gamma_{(\pm)} \equiv  \left\{ \right\} \mp \frac 1 2\,g^{-1}
H\,,\label{cons}
\end{eqnarray}
with $\left\{ \right\}$ the standard Levi-Civita connection. The curvature tensor on ${\cal M}$ calculated with the $\Gamma_{(+)}$ and $\Gamma_{(-)}$ connection resp.~is denoted by $R^{(+)}$ and $R^{(-)}$, they are related by $R^{(+)}_{abcd}=R^{(-)}_{cdab}$. From eq.~(\ref{acnorsp}) it is clear that the physical fields $x^a$ and $\psi_\pm^a$ are supplemented with a set of auxiliary fields $F^a$.

We now investigate under which conditions the action eq.~(\ref{an11}) exhibits an $N=(2,2)$ supersymmetry  
\cite{Gates:1984nk}-\cite{Spindel:1988nh}. The most general expression we can write down for the two additional supersymmetries taking into account super-Poincar\'e invariance and dimensions is given by,
\begin{eqnarray}
\delta x^a= \varepsilon^+J_{+b}^a(x)D_+x^b+ \varepsilon^-J_{-b}^a(x)D_-x^b\,.\label{extrasusy}
\end{eqnarray} 
Calculating the algebra of these transformations yields,
\begin{eqnarray}
[\delta(\varepsilon_1), \delta(\varepsilon_2)]x^a&=& -i \varepsilon _1^+ \varepsilon_2^+(J_+^{\,2}){}^a{}_b \partial _\pp\, x^b
 -i \varepsilon _1^- \varepsilon_2^-(J_-^{\,2}){}^a{}_b \partial _= x^b \nonumber\\
&&-2 \varepsilon_1^+ \varepsilon _2^+ {\cal N}[J_+,J_+]^a{}_{bc}D_+x^bD_+x^c
-2 \varepsilon_1^- \varepsilon _2^- {\cal N}[J_-,J_-]^a{}_{bc}D_-x^bD_-x^c \nonumber\\
&&+ \big(\varepsilon_1^+ \varepsilon _2^-+\varepsilon_1^- \varepsilon _2^+\big)[J_+,J_-]^a{}_b
\big(D_+D_-x^b+\Gamma^b_{(+)dc}D_+x^cD_-x^d\big)+\cdots\,,\label{alg22}
\end{eqnarray}
where we omitted terms linear in $\nabla^{(+)}J_+$ and $\nabla^{(-)}J_-$. We denoted the Nijenhuis tensor\footnote{ Out of two $(1,1)$ tensors $R^a{}_b$ and $ S^a{}_b$,
one constructs a $(1,2)$ tensor ${\cal N}[R,S]^a{}_{bc}$, the Nijenhuis
tensor, as ${\cal N}[R,S]^a{}_{bc}=
R^a{}_dS^d{}_{[b,c]}+R^d{}_{[b}S^a{}_{c],d}+R\leftrightarrow S$. For an almost complex structure $J$, the vanishing of ${\cal N}[J,J]$ is equivalent to
$ [U,V]+J[J U,V]+J[U,J V]- [J U,J V]=0$ with $U$ and $V$ two vectors.} by ${\cal N}$. The non closure terms in the last line of eq.~(\ref{alg22}) are proportional to the equation of motion for $x$ following from the action in eq.~(\ref{an11}) and closure of the algebra is generically only guaranteed on-shell. So one concludes that the transformations given in eq.~(\ref{extrasusy}) satisfy the supersymmetry algebra {\em on-shell} if $J_+$ and $J_-$ are both 
complex structures, {\em i.e.} $J_+^2=J_-^2=-{\bf 1}$ and $ {\cal N}[J_\pm,J_\pm]=0$. In addition the complex structures should be covariantly constant, 
$\nabla^{(+)}J_+=\nabla^{(-)}J_-=0$. 

The action eq.~(\ref{an11}) is invariant under eq.~(\ref{extrasusy}) provided that the metric is hermitian with respect to {\em both} complex structures, $g(J_\pm U,J_\pm V)=g(U,V)$ and, as with the algebra, the complex structures should be covariantly constant. 

Given vectors $U$, $V$ and $W$ we can summarize the conditions under which the non-linear $\sigma$-model possesses an $N=(2,2)$ supersymmetry by:
\begin{itemize}
\item $J_+$ and $J_-$ are two complex structures, {\em i.e.},
\begin{eqnarray}
J_+^2=J_-^2=-{\bf 1}\,,\qquad [U,V]+J_\pm[J_\pm U,V]+J_\pm[U,J_\pm V]- [J_\pm U,J_\pm V]=0\,.
\end{eqnarray} 
\item The metric $g$ is hermitian w.r.t. both complex structures: 
\begin{eqnarray}
g(J_\pm U,J_\pm V)=g(U,V)\,.
\end{eqnarray}
\item The exterior derivative of the two-forms $\omega_\pm(U,V)=-g(U,J_\pm V)$ is given by, 
\begin{eqnarray}
 d\omega _\pm\big(U,V,W\big)=\mp H(J_\pm U,J_\pm V,J_\pm W)\,,\label{cddd}
\end{eqnarray}
\end{itemize}
The last condition is equivalent -- using the previous conditions -- to the requirement that the complex structures are covariantly constant. A target space possessing these properties is often called a bi-hermitian geometry.

The non-closure terms in the algebra eq.~(\ref{alg22}) are proportional to the commutator of the complex structures, $[J_+,J_-]$, so we expect that no auxiliary fields will be needed along $\ker [J_+,J_-]$ while the requirement of off-shell closure  will necessitate the introduction of additional auxiliary fields along $\mbox{im}\big({[}J_+,J_-{]}g^{-1})$. In section 4 we present a full off-shell formulation of the model by reformulating it in $N=(2,2)$ superspace. However we first introduce a geometric formalism in which the above structure finds a natural setting.

\section{Generalized K\"ahler geometry and $N=(2,2)$ non-linear $\sigma$-models}
Because of the presence of the Kalb-Ramond two-form $b$, a coordinate transformation often occurs in combination with a $b$ gauge transformation. {\em E.g.}
a coordinate transformation generated by a vector $X$ acts as $g \rightarrow  g+{\cal L}_Xg$ and $H \rightarrow H+{\cal L}_X H$ on $g$ and $H$. It is an isometry if $g$ and $H$ are invariant. Using that on forms\footnote{We define the interior product so that
the vector is always contracted with the first argument of the form. {\em I.e.} $\iota_X \omega= 
\omega(X,\cdots)\,$.} ${\cal L}_X=d\iota_x+\iota_Xd$ holds, we find ${\cal L}_XH=d {\cal L}_Xb$, so under 
the coordinate transformation we get that $b \rightarrow b+ {\cal L}_Xb+d\xi$ for some one-form $\xi$. Calculating the commutator of 
two such transformations generated by $(X,\xi)$ and $(Y,\eta)$ we find that it closes giving a transformation parameterized 
by $([X,Y],  {\cal L}_X \eta - {\cal L}_Y \xi +\cdots)$ where the dots stands for closed terms.

This phenomenon was captured by Hitchin and Gualtieri in a natural generalization of K\"ahler geometry called generalized K\"ahler Geometry (GKG) \cite{Hitchin:2004ut}. The central object is the bundle $T\oplus T^\ast$, where $T$ and $T^\ast$ are the tangent and the cotangent bundle resp., on which one defines a symmetric bilinear pairing,
\begin{eqnarray}
\langle \IX, \IY\rangle  = \frac 1 2 \left( \iota_X\eta + \iota_Y\xi \right) ,\label{np}
\end{eqnarray}
where $\IX=X+\xi,\,\IY=Y+\eta\,\in T\oplus T^\ast$. This pairing has a large isometry group which includes generalized coordinate transformations and $b$-transformations as subgroups. The latter act as,
\begin{eqnarray}
\IX \rightarrow  e^b \,\IX=\IX+\iota_X b\,,\label{btransfX}
\end{eqnarray}
where $b$ is a locally defined two-form.  In addition this pairing supports a natural action of a bi-vector. 

The Courant bracket replaces the Lie bracket,
\begin{eqnarray}
\Ilb \IX,\IY\Irb = [X,Y] + {\cal L}_X \eta - {\cal L}_Y \xi - \frac 1 2 d \big(\iota_X\eta- \iota_Y\xi\big)\,.\label{Courant}
\end{eqnarray}
While anti-symmetric, the Courant bracket only satisfies the Jacobi identities
on an isotropic subspace\footnote{An isotropic subspace $L \subset T\oplus T^\ast$ is defined by
$\forall \,\IX,\,\IY\in L: \langle\IX,\IY\rangle =0$. If the dimension of $T\oplus T^\ast$ is $2m$, then the maximal dimension of $L$ is given by $m$. Whenever the dimension of $L$ is $m$, we talk about a maximal isotropic subspace. A simple but important example of a maximal isotropic is $T \subset T\oplus T^\ast$. 
} of $T\oplus T^\ast$ provided it acts involutively on that isotropic subspace.  The $b$-transformation eq.~(\ref{btransfX}) acts as ,
\begin{eqnarray}
\Ilb e^b (\IX), e^b (\IY)\Irb = e^b \Ilb\IX,\IY\Irb_{db}\,,
\end{eqnarray}
where $\Ilb \IX,\IY\Irb_{db}=\Ilb \IX,\IY\Irb+\iota_Y\iota_X db$.  Thus the isometries of the of the inner product which hold as symmetries of the Courant bracket consist of $GL(D)$ transformations (evident from the index free formulation) and the action of closed two-forms.

The spinor representation of the isometry group of eq.~(\ref{np}) is realized on the exterior algebra on $T^\ast$. For $\IX=X+\xi\in T\oplus T^\ast$ we introduce a ``gamma matrix'' $ \Gamma_\IX$ which act on
$\phi\in \wedge^\bullet T^\ast$ as,
\begin{eqnarray}
\Gamma_\IX\cdot\phi=\iota_X\phi+\xi\wedge \phi\,.
\end{eqnarray}
A direct check shows that the poly-forms $\wedge^\bullet T^\ast$ do indeed provide a module for the Clifford algebra since:
\begin{eqnarray}
\big\{ \Gamma_\IX, \Gamma_\IY\big\}\cdot\phi=2\langle \IX,\IY\rangle\,\phi\,.\label{dirac}
\end{eqnarray}
Having the gamma matrices, one gets the spin representation of the isometry group of the bilinear form
in the usual way. {\em E.g.} for the $b$ transform, see eq.~(\ref{btransfX}), one gets,
\begin{eqnarray}
\phi \rightarrow e^{-b\wedge}\,\phi\,.\label{bspin}
\end{eqnarray}
Similarly a spinor transforms under a coordinate transformation as a density,
\begin{eqnarray}
x \rightarrow x'(x)\Rightarrow \phi(x) \rightarrow \phi'(x')=\sqrt{\det \frac{\partial x'}{\partial x}}\, 
\phi(x)\,.
\end{eqnarray}
In fact this transformation law implies the isomorphism between spinors and poly-forms is actually given by 
\begin{eqnarray} \label{spinoriso}
S \cong (\det T)^{\frac{1}{2}} \wedge^\bullet T^\ast \ . 
\end{eqnarray}
The Mukai pairing gives an invariant (under the isometry group of $\langle\cdot,
\cdot\rangle$ connected to the identity) bilinear for the spinors,
\begin{eqnarray}
\big( \phi_1,\phi_2\big)= \sigma( \phi_1)\wedge \phi_2|_{\mbox{top}}\,,
\end{eqnarray}
where $\phi_1,\, \phi_2\in\wedge^\bullet T^\ast$ and $\sigma$ acts as,
\begin{eqnarray}
\sigma\big(dx^1\wedge dx^2\wedge \cdots\wedge dx^r\big)=
dx^r\wedge dx^{r-1}\wedge \cdots\wedge dx^1\,.
\end{eqnarray}
That this bilinear produces a top-form rather than a scalar is reflected by the isomorphism eq.~(\ref{spinoriso}).

There is a natural way to associate an isotropic subspace $L$ of $T\oplus T^\ast$ to a given spinor 
$\phi$:
\begin{eqnarray}
\IX\in L \Leftrightarrow \Gamma_\IX\cdot\phi=0\,.
\end{eqnarray} 
Using eq.~(\ref{dirac}) one immediately shows that $L$ is indeed isotropic. When $L$ is maximally 
isotropic one calls $\phi$ a {\em pure spinor}.   As shown by Gualtieri, any non-degenerate pure spinor can be written in the form,
\begin{eqnarray}
\phi =\kappa\wedge e^{i \Omega +\Xi}\,,
\end{eqnarray}
where $\Omega$ and $\Xi $ are real two-forms and $\kappa$ is a complex decomposable $k$-form where $k$ is the type of the generalized complex structure.

A {\em generalized complex structure} (GCS) is defined as a
linear map ${\cal J}:  T \oplus T^\ast \rightarrow T \oplus T^\ast$,
satisfying ${\cal J}^2 = -1$, for which the natural pairing is ``hermitian'',
$\langle {\cal J} \IX, {\cal J} \IY \rangle = \langle \IX, \IY \rangle$
for all $\IX, \IY \in T \oplus T^\ast$ and for which the $+i$ eigenbundle is involutive under the 
Courant bracket,
\begin{eqnarray}
 \Ilb \IX,\IY \Irb +{\cal J}\Ilb {\cal J} \IX,\IY\Irb +{\cal J}\Ilb \IX,{\cal J} \IY\Irb -\Ilb{\cal J} 
\IX,{\cal J} \IY\Irb =0.
\end{eqnarray}
In order for a GCS to exist the manifold $\cal M$ should be even dimensional and we take
$\dim {\cal M}=2m$.

Note that if ${\cal J}$ is integrable w.r.t. the bracket $\Ilb\cdot,\cdot\Irb$ then $e^{-b} {\cal J} e^{b}$ is
integrable w.r.t. the bracket $\Ilb\cdot,\cdot\Irb_{db}$. Writing $(T\oplus T^\ast)\otimes \IC=L\oplus \bar L$ where $L$ ($\bar L$) is the
$+i$ ($-i$) eigenbundle of $ {\cal J}$, we see that $L$ is a maximal isotropic subspace
of $(T\oplus T^\ast)\otimes \IC$. As a consequence, any GCS comes with a pure spinor $\phi$ defined
by $ \Gamma_\IX\cdot\phi=0$, $\forall\, \IX\in L$. For a physicist this is just the highest weight
vector of the spinor representation. 

Both a regular complex structure $J$ and a  symplectic structure (a closed non-degenerate two-form) $\omega$ give rise to a GCS, 
\begin{eqnarray}
{\cal J}_c =  \left( \begin{array}{cc}
                                        J & 0 \\
                                        0 & -J^t
                                        \end{array}\right)\,,
\qquad                                        
{\cal J}_s = \left( \begin{array}{cc}
                                        0 & \omega^{-1} \\
                                        -\omega & 0
                                        \end{array}\right)\,.\label{canpusp}
\end{eqnarray}
A generic GCS interpolates between these two extreme cases. Gualtieri extended the Newlander-Nirenberg (for complex structures) and the Darboux theorem (for symplectic structures) to an arbitrary
GCS: by an appropriate diffeomorphism and $b$-transformation one can 
always turn a GCS to the standard product GCS $\IC^k\times (\IR^{2m-2k}, \omega)$. The integer 
$k$ is called the {\em type} of the GCS. 
So a $2m$ dimensional manifold with a generalized 
complex structure is foliated by $2m-2k$ dimensional leaves of the form 
$\IR^{2m-2k} \times  \{ {\rm 
point}\}$ on which a symplectic form $\omega$ is properly defined. Transverse to the leaves, 
complex coordinates $z_i$ with $i \in \{ 1, \ldots , k\}$ are introduced such that the leaves are 
located at $z_i = {\rm constant} \, (\forall \, i)$. A generic feature of generalized complex geometry is that loci might exist where the 
type jumps, one calls this phenomenon {\em type changing}. Gualtieri's theorem  holds for neighborhoods of regular points ({\em i.e.} where no type changing occurs). 

A {\em generalized  K{\"a}hler structure} (GKS) requires two mutually
commuting GCS's, ${\cal J}_+$ and ${\cal J_-}$ such that,
\begin{eqnarray}
{\cal G}\big(\IX,\IY\big)=\langle {\cal J}_+\IX, {\cal J}_- \IY\rangle,
\end{eqnarray}
defines a positive definite metric on $T\oplus
T^*$. Gualtieri showed that a GKS is exactly equivalent to the previously introduced $N=(2,2)$ non-linear $\sigma $-model gemometry (see section 2),
\begin{eqnarray}
{\cal J}_{\pm} = \frac 1 2\left( \begin{array}{cc}1&0\\-b&1 \end{array}\right)\,
                                        \left( \begin{array}{cc}
                                        J_+ \pm J_- & \omega^{-1}_+ \mp \omega^{-1}_- \\
                                        -(\omega_+ \mp \omega_-) & -(J^t_+ \pm J^t_-)
                                        \end{array}\right)\,
                                        \left( \begin{array}{cc}1&0\\+b&1 \end{array}\right)\,.
                                        \label{gcs22}
\end{eqnarray}

We call the types of the generalized complex structures ${\cal J}_+$
and
${\cal J}_-$ , $k_+$ and $k_-$ resp. From 
eq.~(\ref{gcs22})  we get that   
\begin{eqnarray}
\big(k_+,k_-\big)=\frac 1 2 \,\big(\dim\ker(J_+-J_-),\dim\ker(J_++J_-)\big).
\end{eqnarray}
As $ {\cal J_+}$ and $ {\cal J_-}$ commute, we can write,
\begin{eqnarray}
\big(T\oplus T^\ast\big)\otimes \IC=
L_{++}\oplus L_{+-}\oplus L_{--}\oplus L_{-+},
\end{eqnarray}
where $L_{++}$ is the $+i$ eigenbundle for both $ {\cal J_+}$ and $ {\cal J_-}$; 
$L_{+-}$ is the $+i$ eigenbundle for $ {\cal J_+}$ and the $-i$ eigenbundle for $ {\cal J_-}$; etc.
Denoting $\IX_+\in L_{++}$, $\IX_-\in L_{+-}$, $\bar\IX_+\in L_{--}$ and $\bar\IX_-\in L_{-+}$, we can introduce pure spinors $\phi_+$ and $\phi_-$ for $ {\cal J_+}$ and $ {\cal J_-}$ resp. defined by,
\begin{eqnarray}
\Gamma_{\IX_+}\cdot\phi_+=\Gamma_{\IX_-}\cdot\phi_+=0,\qquad
\Gamma_{\IX_+}\cdot\phi_-=\Gamma_{\bar\IX_-}\cdot\phi_-=0.\label{psdef}
\end{eqnarray}
Eq.~(\ref{psdef}) does not fix the normalization of the pure spinors $\phi_+$ and $\phi_-$. 
It can be shown \cite{Hitchin:2004ut} that the integrability of the generalized complex structures 
guarantees the existence of $\IY_+$ and $\IY_-$ such that,
\begin{eqnarray}
d \phi_\pm= \Gamma_{\IY_\pm}\cdot \phi_\pm\,.
\end{eqnarray}

Finally, a {\em generalized Calabi-Yau geometry} is a generalized K\"ahler geometry for which the
pure spinors $\phi_+$ and $\phi_-$ are {\em globally defined}, {\em closed} and they satisfy,
\begin{eqnarray}
\big(\phi_+,\bar \phi_+\big)= c\, \big(\phi_-,\bar \phi_-\big)\neq 0\,,\label{gcycd}
\end{eqnarray}
with $c$ constant.

\section{The off-shell realization of $N=(2,2)$ supersymmetry}
We now pass to $N=(2,2)$ superspace. Because of dimensional reasons the Lagrange density for the $N=(2,2)$ non-linear $\sigma$-model will be  a function of scalar superfields so all dynamics will arise from the constraints satisfied by the superfields. We will only consider constraints linear in the derivatives. The maximal set of constraints we can impose on a set of scalar superfields $x^a$ consistent with dimensions and super-Lorentz invariance is \cite{Maes:2006bm},
\begin{eqnarray}
\hat D_+x^a= J^a_+{}_b(x)D_+x^b\,,\qquad \hat D_-x^a= J^a_-{}_b(x)D_-x^b\,.\label{cccsf}
\end{eqnarray}
Eqs.~(\ref{app4}) then imply integrability conditions which state that $J_+$ and $J_-$ are complex structures, {\em i.e.} $J_+^2=J_-^2=-{\bf 1}$ and ${\cal N }[J_\pm,J_\pm]=0$.
From $\{\hat D_+,\hat D_-\}=0$ follows that $J_+$ and $J_-$ commute, $[J_+,J_-]=0$ and that  $ {\cal M}[J_+,J_-]$ vanishes\footnote{
Out of two {\em commuting} $(1,1)$ tensors $R$ and $S$  one constructs the tensor  ${\cal M}[R,S]$,
${\cal M}[R,S](U,V)=
[RU,SV]-R[U,SV]-S[RU,V]+RS[U,V]$. It is related to the Nijenhuis tensor by ${\cal N}[R,S](U,V)=\frac 1 2 \left( {\cal M}[R,S](U,V) + {\cal M}[S,R](U,V)\right)$.}.

So $J_+$ and $J_-$ are mutual commuting complex structures which can be simultaneously diagonalized. This gives rise to two cases: either both $J_+$ and $J_-$ have the same eigenvalue or they have opposite eigenvalues. In the first case we call them {\bf chiral superfields}, in a complex basis they satisfy,
\begin{eqnarray}
\bar\ID_+ z=\bar\ID_- z=0\,,\qquad \ID_+\bar z=\ID_-\bar z=0\,.
\end{eqnarray}
The latter case gives rise to {\bf twisted chiral superfields} \cite{Gates:1984nk} satisfying,
\begin{eqnarray}
\bar\ID_+ w=\ID_- w=0\,,\qquad \ID_+\bar w=\bar \ID_-\bar w=0\,.
\end{eqnarray}
The only possibility left is constraining a single chirality. In order to get a $\sigma$-model action they need to come in pairs satisfying constraints of opposite chirality. They are called {\bf semi-chiral superfields} \cite{Buscher:1987uw} and they are defined by,
\begin{eqnarray}
\bar\ID_+l=\bar\ID_-r=0,\qquad \ID_+\bar l=\ID_-\bar r=0,\qquad
\end{eqnarray}
An unconstrained $N=(2,2)$ superfield consists of four independent $N=(1,1)$ superfields. When dealing with chiral or twisted chiral superfields, the constraints reduce the number of components of an $N=(2,2)$ superfield to those of an $N=(1,1)$ superfield. A semi-chiral superfield describes twice as many degrees of freedom compared to an $N=(1,1)$ superfield, half of which will turn out to be auxiliary. 

Decomposing the tangent space as $T=\ker(J_+-J_-)\oplus\ker(J_++J_-)\oplus\mbox{im}\big({[}J_+,J_-{]}g^{-1}\big)$, one can show that $\ker(J_--J_-)$, $\ker(J_++J_-)$ and $\mbox{im}\big({[}J_+,J_-{]}g^{-1})$ can be integrated to chiral, twisted chiral and semi-chiral fields resp. (this was
conjectured in \cite{Sevrin:1996jr} and proven in \cite{Lindstrom:2005zr}).

The most general action involving these superfields in $N=(2,2)$ superspace is given by,
\begin{eqnarray}
{\cal S}=4\,\int\,d^2 \sigma \,d^2\theta \,d^2 \hat \theta \, V(l,\bar l,r,\bar r,w,\bar w, z, \bar z),
\label{actionN22}
\end{eqnarray}
where the Lagrange density $V(l,\bar l,r,\bar r,w,\bar w, z, \bar z)$ is
an arbitrary real function of the semi-chiral,  $l^{\tilde \alpha }$, $l^{\bar{ \tilde
  \alpha}  }$, $r^{\tilde \mu  }$, $r^{\bar{ \tilde \mu }  }$,
  $\tilde \alpha,\,\bar{ \tilde \alpha} ,\,\tilde \mu ,\,\bar{
  \tilde \mu }\in \{1,\cdots n_s\}$, the twisted chiral,
 $w^\mu $, $w^{\bar \mu }$, $\mu,\,\bar \mu
  \in\{1,\cdots n_t\}$, and the
chiral superfields, 
$z^\alpha $, $z^{\bar \alpha }$, $\alpha,\, \bar \alpha
  \in\{1,\cdots n_c\}$. It is defined modulo a generalized K{\"a}hler
transformation,
\begin{eqnarray}
V\rightarrow V+F(l,w,z)+ \bar F(\bar l,\bar w,\bar z)+ G(\bar r,w,\bar z)+ \bar G(r,\bar w,z).\label{genKahltrsf1}
\end{eqnarray}
As for the usual K{\"a}hler case these generalized K{\"a}hler transformations
are essential for the global consistency of the model, see {\em e.g.} \cite{Hull:2008vw}. Finally, there is the
local mirror transformation which sends $J_-$ to $-J_-$ which at the level of the generalized K\"ahler
potential amounts to,
\begin{eqnarray}
V(l,\bar l,r,\bar r,w,\bar w, z, \bar z) \rightarrow 
V(l,\bar l,\bar r, r,z,\bar z, w, \bar w)\,.\label{mirror}
\end{eqnarray}

Before proceeding, we introduce some
notation. We write,
\begin{eqnarray}
 M_{AB}=\left( \begin{array}{cc}
   V_{ab} & V_{a\bar b} \\
   V_{\bar a b} & V_{\bar a\bar b}
 \end{array}\right),
\end{eqnarray}
where, $(A,a)\in\{(l,\tilde \alpha ),\,(r,\tilde \mu ),\,(w,\mu
),\,(z,\alpha )\}$ and $(B,b)\in\{(l,\tilde \beta ),\,(r,\tilde \nu
),\,(w,\nu ),\,(z,\beta )\}$. The subindices on $V$ denote derivatives
with respect to those coordinates. In this way {\em e.g.} we get that
$M_{wl}$ is the $2n_t\times 2n_s$ matrix given by,
\begin{eqnarray}
 M_{wl}=\left( \begin{array}{cc}
   V_{\mu \tilde \beta } & V_{\mu \bar{\tilde \beta }} \\
   V_{\bar \mu \tilde \beta } & V_{\bar \mu \bar{\tilde \beta }}
 \end{array}\right).
\end{eqnarray}
We also introduce the matrix $\IJ$,
\begin{eqnarray}
 \IJ\equiv i\,\left(
\begin{array}{cc}
  {\bf 1} & 0 \\
  0 & -{\bf 1}
\end{array}
 \right),
\end{eqnarray}
with $\bf 1$ the unit matrix and using this we write,
\begin{eqnarray}
 C_{AB}\equiv \IJ\, M_{AB}-M_{AB}\,\IJ,\qquad
A_{AB}\equiv \IJ\, M_{AB}+M_{AB}\,\IJ. \label{canda}
\end{eqnarray}

Starting from the action eq.~(\ref{actionN22}) one passes to $N=(1,1)$ superspace and eliminates the
auxiliary fields. Comparing this to eqs. (\ref{cccsf}) and  (\ref{an11}) one gets explicit expressions for the complex structures, the metric and the Kalb-Ramond two-form. The complex structures are given by,
\begin{eqnarray}
 J_+&=&\left(\begin{array}{cccc}
   \IJ & 0 & 0 & 0 \\
   M_{lr}^{-1}C_{ll} &  M_{lr}^{-1}\IJ M_{lr} &
   M_{lr}^{-1}C_{lw}&M_{lr}^{-1} C_{lz}   \\
   0 & 0 & \IJ & 0 \\
   0 & 0 & 0 & \IJ
 \end{array}   \right),\nonumber\\
J_-&=&\left(\begin{array}{cccc}
   M_{rl}^{-1}\IJ M_{rl} &  M_{rl}^{-1} C_{rr} &
   M_{rl}^{-1}A_{rw}&M_{rl}^{-1} C_{rz}    \\
0&\IJ  & 0 & 0 \\
   0 & 0 & -\IJ & 0 \\
   0 & 0 & 0 & \IJ
 \end{array}   \right),\nonumber\\
\end{eqnarray}
where we labeled rows and columns in the order $l$, $\bar l$, $r$, $\bar
r$, $w$, $\bar w$, $z$, $\bar z$. Note that once semi-chiral fields are
present, neither of the complex structures is diagonal. One easily shows
\cite{Sevrin:1996jr} that making a coordinate transformation,
\begin{eqnarray}
r^{\tilde \mu }\rightarrow V_{\tilde \alpha }\,,\qquad r^{\bar {\tilde \mu} } \rightarrow V_{\bar{\tilde \alpha} }\,,
\end{eqnarray}
while keeping the other coordinates unchanged diagonalizes $J_+$. Similarly, the coordinate
transformation,
\begin{eqnarray}
l^{\tilde \alpha }\rightarrow V_{\tilde \mu }\,,\qquad l^{\bar {\tilde \alpha} } \rightarrow V_{\bar{\tilde \mu} }\,,
\end{eqnarray}
with the other coordinates fixed diagonalizes $J_-$. This observation led to a reinterpretation in \cite{Lindstrom:2005zr} of the generalized K\"ahler potential as the generating function for a canonical transformation which interpolates between the
coordinate system in which $J_+$ is diagonal and the one in which $J_-$
is diagonal.

{}From the second order action one reads off the metric $g$ and the
torsion two-form potential $b$. One finds two natural expressions,
\begin{eqnarray}
&&\big(g-b_+\big)(X,Y)=\Omega ^+(X,J_+Y)=d B^+\,(X,J_+Y)\,, \nonumber\\
&&\big(g+b_-\big)(X,Y)=\Omega ^-(X,J_-Y)=dB^-\,(X,J_-Y)\,.\label{gminb}
\end{eqnarray}
where $\Omega^+$ and $\Omega^-$ are two (locally defined) closed two-forms linear in the
generalized K{\"a}hler potential and $H=db_+=db_-$. From eq.~(\ref{gminb}) one gets that $b_\pm$
are $(2,0)$+$(0,2)$ forms w.r.t. $J_\pm$.  
Explicitly we get,
\begin{eqnarray}
\Omega^+ = -\frac 1 2 \left(\begin{array}{cccc} C_{ll} & A_{lr} &
C_{lw} & A_{lz}\\
-A_{rl} & -C_{rr} & -A_{rw} & -C_{rz} \\
C_{wl}& A_{wr} & C_{ww} & A_{wz} \\
-A_{zl} & -C_{zr} & -A_{zw} & -C_{zz}
\end{array}
 \right), \qquad 
\Omega^- = \frac 1 2 \left(\begin{array}{cccc} C_{ll} & C_{lr} &
C_{lw} & C_{lz}\\
C_{rl} & C_{rr} & C_{rw} & C_{rz} \\
C_{wl}& C_{wr} & C_{ww} & C_{wz} \\
C_{zl} & C_{zr} & C_{zw} & C_{zz}
\end{array}
 \right).\label{ommin}
\end{eqnarray}
where we labeled rows and columns in the order $(l, \bar l, r, \bar r,
w, \bar w,z,\bar z)$ and locally we get $\Omega^\pm=dB^\pm$ where\footnote{The notation
we use is such that $V_l\,dl$ stands for $\partial_{l^{\tilde{ \alpha}}}  V\,dl^{\tilde{ \alpha}}\,$, $V_{\bar l}\,d\bar l$ for $\partial_{l^{\bar{\tilde{ \alpha}}}}  V\,dl^{\bar{\tilde{ \alpha}}}\,$, etc.},
\begin{eqnarray}
2 \,B^+ &=&  i\, V_l\,dl-i\,V_{\bar l}\,d\bar l - i\,V_r\,dr+i\, V_{\bar r}\,d\bar r+i\,V_w\,dw -
i\,V_{\bar w}\,d\bar w -i\,V_z\,dz +i V_{\bar z}\,d\bar z \,,\nonumber\\
 2 \,B^- &=&  -i\, V_l\,dl+i\,V_{\bar l}\,d\bar l - i\,V_r\,dr+i\, V_{\bar r}\,d\bar r
 -i\,V_w\,dw + i\,V_{\bar w}\,d\bar w -i\,V_z\,dz +i V_{\bar z}\,d\bar z\, .
\end{eqnarray}
There is a certain freedom in the definition of $ \Omega^\pm$,
\begin{eqnarray}
\Omega^\pm(X,Y)\simeq \Omega^\pm(X,Y)+d\xi^\pm(X,J_\pm Y)\,,
\end{eqnarray}
where $d\xi^+$ and $d\xi^-$ are $(2,0)+(0,2)$ forms w.r.t. $J_+$ and $J_-$ resp. Examples of this 
are $\xi^+=V_l\,dl+V_{\bar l}\, d\bar l$,  $\xi^+=i\,V_l\,dl-i\,V_{\bar l}\, d\bar l$,  
$\xi^-=V_r\,dr+V_{\bar r}\, d\bar r$ and  $\xi^-=i\,V_r\,dr-i\,V_{\bar r}\, d\bar r$. For the choice in  eq.~(\ref{ommin}) one gets that when no chiral
fields are present we can write $ \Omega^+$ as \cite{Bogaerts:1999jc},
\begin{eqnarray}
  \Omega^+ (X,Y)= 2\,g\big(X, (J_+-J_-)^{-1}Y\big).\label{opglob}
\end{eqnarray}
Similarly, when there are no twisted chiral fields we get,
\begin{eqnarray}
  \Omega^- (X,Y)= 2\,g\big(X, (J_++J_-)^{-1}Y\big).\label{omglob}
\end{eqnarray}
Note that these expressions only exist in regular neighborhoods, at loci where type changing occurs they might not exist.
Using the previous expressions one also finds,
\begin{eqnarray}
b_--b_+=\frac 1 2 \,d\big(
-V_l\,dl-V_{\bar l}\,d\bar l
+V_r\,dr+V_{\bar r}\,d\bar r
-V_w\,dw-V_{\bar w}\,d\bar w
+V_z\,dz+V_{\bar z}\,d\bar z
\big).
\end{eqnarray}

In \cite{Grisaru:1997pg} the 1-loop $\beta$-function for a general $N=(2,2)$ $\sigma$-model in superspace was calculated. The counter term 
which was found reads,
\begin{eqnarray}
{\cal S}_{1-\mbox{loop}}\,\sim\frac{1}{\varepsilon}\,\int d^2 \sigma \,d^2 \theta \, d^2 \hat \theta \,\ln\,\frac{\det\big(N_+\big)}{\det\big(N_-\big)}\,,
\end{eqnarray}
where,
\begin{eqnarray}
N_+=\left(
\begin{array}{ccc}
V_{l\bar l}&V_{lr}&V_{l\bar w}\\
V_{\bar r\bar l}&V_{\bar r r}&V_{\bar r \bar w}\\
V_{w\bar l}&V_{wr}&V_{w\bar w}\end{array}
\right),
\end{eqnarray}
and,
\begin{eqnarray}
N_-=\left(
\begin{array}{ccc}
V_{l\bar l}&V_{l\bar r}&V_{l\bar z}\\
V_{r\bar l}&V_{r\bar r }&V_{ r \bar z}\\
V_{z\bar l}&V_{z\bar r}&V_{z\bar z}\end{array}
\right)\,.
\end{eqnarray}
It vanishes if,
\begin{eqnarray}
\frac{\det\big(N_+\big)}{\det\big(N_-\big)}\,=\pm|f_+(l,w,z)|^2
|f_-(r, \bar{w},z)|^2\,
,\label{CYcd2}
\end{eqnarray}
for some functions $f_+$ and $f_-$.

Let us now return to the generalized K\"ahler geometry.  The closed pure spinors, eqs.~(\ref{psdef}), are given by \cite{Hull:2010sn}, \cite{Sevrin:2011mc},
\begin{eqnarray}
\phi_+&=&d\bar z^1\wedge d\bar z^2\wedge \cdots\wedge  d\bar z^{n_c}\wedge
e^{i\, \Omega^++\Xi^+}\,, \nonumber\\
\phi_-&=&d\bar w^1\wedge d\bar w^2\wedge \cdots\wedge  d\bar w^{n_t}\wedge
e^{i\, \Omega^-+\Xi^-}\,,\label{psexpl}
\end{eqnarray} 
where $\Omega^\pm$ are given in eq.~(\ref{ommin}) and $\Xi^\pm$ are
given by,
\begin{eqnarray}
\Xi^+&=&\frac 1 2 \,d\big(V_l\,dl+V_{\bar l}\,d\bar l
-V_z\,dz-V_{\bar z}\,d\bar z\big)\,, \nonumber\\
\Xi^-&=&\frac 1 2 \,d\big(V_r\,dr+V_{\bar r}\,d\bar r
-V_w\,dw-V_{\bar w}\,d\bar w\big)\, .\label{xidef}
\end{eqnarray} 
Using this one gets the Mukai pairings,
\begin{eqnarray}
\big(\phi_+,\bar \phi_+\big)&=&(-1)^{n_c(n_c+1)/2+n_t+n_s}\,2^{n_t+2n_s}\,\det N_+\, \nonumber\\
\big(\phi_-,\bar \phi_-\big)&=&(-1)^{n_t(n_t+1)/2}\,2^{n_c+2n_s}\,\det N_-\, ,
\label{mukaip}
\end{eqnarray}
from which one gets  the generalized Calabi-Yau condition  eq.~(\ref{gcycd}),
\begin{eqnarray}
\frac{\det\big(N_+\big)}{\det\big(N_-\big)}\,=\mbox{ constant }\neq 0\,.\label{CYcd1}
\end{eqnarray}
It is clear that eq.~(\ref{CYcd1}) is stronger than eq.~(\ref{CYcd2}) which reflects the fact that the vanishing of the $\beta$-functions is necessary but not sufficient for the background to solve the supergravity equations of motion. For a more detailed discussion see {\em e.g.} \cite{Hull:2010sn} and references therein.

\section{Duality transformations}
In $N=(2,2)$ supersymmetric models there exists a variety of duality
transformations which allows one to change the nature of the superfields. 
A complete catalogue of duality transformations in $N=(2,2)$
superspace was obtained in \cite{Grisaru:1997ep}, the extension of this to the case where boundaries are present was developed in \cite{Sevrin:2009na}.
A first class of duality transformations is always possible and is a
consequence of the constraints satisfied by $N=(2,2)$
superfields. One imposes the
constraints on the superfields through Lagrange multipliers
(unconstrained superfields). In such a first order formulation the
original fields are then unconstrained superfields. Integrating over the
Lagrange multipliers results in the original model; 
integrating over the original unconstrained fields yields the dual
formulation. In this way a (twisted) chiral field is dual to a (twisted) complex linear superfield, so the dual models
involve superfields satisfying constraints which are quadratic in the derivatives, a case we do not consider any further here. More interesting are semi-chiral fields where one has a total of 4 dual formulations. 

The starting point is the first order action,
\begin{eqnarray}
{\cal S}&=&-\int\,d^2 \sigma \,d^2\theta \,d^2 \hat \theta \, \Big(V(l,\bar l,r,\bar r,\cdots)-
\Lambda ^ +\,\bar \ID_+ l-\bar \Lambda ^+\,\ID_+ \bar l- \Lambda ^-\,\bar \ID_- r-
\bar \Lambda ^-\,\ID_- \bar r\Big)\,,
\end{eqnarray}
where $l$, $\bar l$, $r$ and $\bar r$ are unconstrained bosonic complex
superfields and $\Lambda ^\pm$ and $\bar \Lambda ^\pm$ are unconstrained
complex fermionic superfields. Integrating over the Lagrange multipliers
constrains $l$ and $r$ to form a semi-chiral multiplet. Upon partial
integration we can rewrite the action in three ways,
\begin{eqnarray}
{\cal S}&=&-\int\,d^2 \sigma \,d^2\theta \,d^2 \hat \theta \, \Big(V(l,\bar l,r,\bar r,\cdots)-
l\, l'-\bar l\, \bar l'- \Lambda ^-\,\bar \ID_- r- \bar \Lambda ^-\,\ID_- \bar r\Big)\nonumber\\
&=&-\int\,d^2 \sigma \,d^2\theta \,d^2  \hat\theta \, \Big(V(l,\bar l,r,\bar r,\cdots)-
\Lambda ^ +\,\bar \ID_+ l-\bar \Lambda ^+\,\ID_+ \bar l- r\, r'- \bar r\, \bar r'\Big)\nonumber\\
&=&-\int\,d^2 \sigma \,d^2\theta \,d^2 \hat \theta \, \Big(V(l,\bar l,r,\bar r,\cdots)-
l\, l'-\bar l\, \bar l'- r\, r'- \bar r\, \bar r'\Big)\,,
\end{eqnarray}
where we introduced the notation $l'=\bar \ID_+\Lambda ^+$, $\bar l'=
\ID_+\bar \Lambda ^+$, $r'=\bar \ID_-\Lambda ^-$, $\bar r'= \ID_-\bar
\Lambda ^-$. Integrating over the unconstrained fields $( l,\bar l
,\Lambda ^-,\bar \Lambda ^-)$, $(\Lambda ^+,\bar \Lambda ^+,r,\bar r)$ or
$(l,\bar l,r ,\bar r )$ resp.~yields three dual formulations of the
model. This is perfectly consistent with the interpretation of the generalized K\"ahler potential as the generating function for a canonical transformation interpolating between coordinates where $J_+$ or $J_-$ is diagonal as the generating function of a canonical transformation also exists in four versions connected by Legendre transformations. 

In fact when considering the global picture of an $N=(2,2)$ model one should allow on the overlap of two coordinate patches not only for 
a generalized K\"ahler transformation eq.~(\ref{genKahltrsf1}) but for Legendre transformations of the potential as well. We will see an example of this in section 6. 

A second class of dualities consists of T-dualities where the $N=(2,2)$ supersymmetry is kept manifest. 
These dualities require the existence of an isometry and the idea is to gauge the isometry while -- using Lagrange
multipliers -- enforcing the gauge fields to be pure gauge. Integrating over
the Lagrange multipliers returns us to the original model while
integrating over the gauge fields yields the dual model. There are three main cases.

\subsection{The duality between a pair of chiral and twisted chiral
fields and a semi-chiral multiplet}

A first class of T-duality transformations exchanges a pair of chiral and twisted chiral superfields for a semi-chiral multiplet or vice-versa\footnote{While this duality transformation
was already found in \cite{Grisaru:1997ep}, the elucidation of the
underlying gauge structure was done in 
\cite{Lindstrom:2007vc}.}. The starting point is a potential of the form $V\big(z+\bar z,w+\bar
w,i(z-\bar z-w+\bar w),\cdots\big)$. This clearly
exhibits the isometry $z\rightarrow z+i\,a$, $w\rightarrow w+i\,a$, with
$a$ an arbitrary real constant. The first order action
is,
\begin{eqnarray}
 {\cal S}&=&-\int\,d^2 \sigma \,d^2\theta \,d^2 \hat \theta\, \Big(V\big(Y,\tilde Y,\hat Y,\cdots\big)
 +\Lambda ^+\bar \ID_+\big(Y-\tilde Y-i\,\hat Y\big)
 +\bar \Lambda ^+ \ID_+\big(Y-\tilde Y+i\,\hat Y\big)\nonumber\\
&&-\Lambda ^-\bar \ID_-\big(Y+\tilde Y-i\,\hat Y\big)
 -\bar \Lambda ^- \ID_-\big(Y+\tilde Y+i\,\hat Y\big)\Big)
\end{eqnarray}
where $\Lambda ^\pm$ and $\bar \Lambda ^\pm$ are unconstrained complex
fermionic superfields and $Y$, $\tilde Y$ and $\hat Y$ are unconstrained
real bosonic superfields. Integrating over the Lagrange multipliers
$\Lambda ^\pm$ and $\bar \Lambda ^\pm$ returns us to the original model.
Integrating by parts results in,
\begin{eqnarray}
 {\cal S}&=&-\int\,d^2 \sigma \,d^2\theta \,d^2  \hat\theta \, \Big(V\big(Y,\tilde Y,\hat Y,\cdots\big)
 +Y\big(l+\bar l-r-\bar r\big)
-\tilde Y\big(l+\bar l+r+\bar r\big)\nonumber\\
&&-i\,\hat Y\big(l-\bar l-r+\bar r\big)\Big)
 \label{1storderCTS}
\end{eqnarray}
where we introduced the semi-chiral multiplet $l=\bar \ID_+\Lambda ^+$,
$\bar l=\ID_+\bar \Lambda ^+$, $r=\bar \ID_-\Lambda ^-$ and $\bar r=
\ID_-\bar \Lambda ^-$. Integrating over $Y$, $\tilde Y$ and $\hat Y$
yields the dual model.

The inverse transformation starts from a potential of the form $V\big(l+\bar l,r+\bar r,i(l-\bar l-r+\bar
r),\cdots\big)$. The basic relations are given by,
\begin{eqnarray}
 {\cal S}&=&-\int\,d^2 \sigma \,d^2\theta \,d^2 \hat \theta \, \Big(V\big(Y,\tilde Y,\hat Y,\cdots\big)
+i\,u\bar \ID_+\bar \ID_-\big(Y-\tilde Y-i\hat Y\big)\nonumber\\
&&+i\,\bar u \ID_+ \ID_-\big(Y-\tilde Y+i\hat Y\big)
-i\,v\bar \ID_+ \ID_-\big(Y+\tilde Y-i\hat Y\big)
-i\,\bar v\ID_+\bar \ID_-\big(Y+\tilde Y+i\hat Y\big)
 \Big)\nonumber\\
  &=&-\int\,d^2 \sigma \,d^2\theta \,d^2  \hat\theta \, \Big(V\big(Y,\tilde Y,\hat Y,\cdots\big)
+Y\big(z+\bar z-w-\bar w\big)-\tilde Y\big(z+\bar z+w+\bar w\big)\nonumber\\
&&-i\,\hat Y\big(z-\bar z-w+\bar w\big)\Big)
 \label{EXb10}
\end{eqnarray}
where $u,\,v\in\IC$ and $Y,\,\tilde Y,\,\hat Y\in\IR$ are unconstrained
superfields and where we defined $z=i\bar \ID_+\bar \ID_-u$, $\bar z=i
\ID_+ \ID_-\bar u$, $w=i\bar \ID_+ \ID_-v$ and $\bar w=i \ID_+\bar
\ID_-\bar v$. 

\subsection{The duality between a chiral and a twisted chiral field}
The duality transformation between a chiral and a twisted chiral superfield was proposed in
\cite{Gates:1984nk}.
Starting from a potential of the form $V(z+\bar z,\cdots)$ we write the first order action,
\begin{eqnarray}
 {\cal S}&=&-\int\,d^2 \sigma \,d^2\theta \,d^2 \hat \theta \, \Big(V\big(Y,\cdots\big)
 -i\,u\,\bar \ID_+\ID_-Y-i\,\bar u \,\ID_+\bar \ID_-Y\Big).\label{xxfoa1}
\end{eqnarray}
Integrating over the complex unconstrained Lagrange multipliers $u$ and
$\bar u$ brings us back to the original model. Upon integrating by parts
one gets,
\begin{eqnarray}
 {\cal S}&=&-\int\,d^2 \sigma \,d^2\theta \,d^2  \theta' \, \Big(V\big(Y,\cdots\big)
 -Y\big(w+\bar w\big)\Big),\label{xxfoa2}
\end{eqnarray}
where we introduced the twisted chiral fields $w=i\,\bar \ID_+\ID_-u$ and
$\bar w=i\,\ID_+\bar \ID_-\bar u$. Integrating over the unconstrained
gauge field $Y$ gives  the dual model in terms of a twisted chiral
field $w$.

The inverse transformation starts from potentials of the form $V(w+\bar
w,\cdots)$. One has
\begin{eqnarray}
&&-\int\,d^2 \sigma \,d^2\theta \,d^2 \hat \theta \, \Big(V\big(\tilde Y,\cdots\big)
 -i\,u\,\bar \ID_+\bar \ID_-\tilde Y-i\,\bar u \,\ID_+\ID_-\tilde Y\Big)\nonumber\\
 &&=-\int\,d^2 \sigma \,d^2\theta \,d^2  \hat\theta \, \Big(V\big(\tilde Y,\cdots\big)
 -\tilde Y\big(z+\bar z)\Big)\,,\label{xxfoa3}
\end{eqnarray}
where we have put $z=i\,\bar \ID_+\bar \ID_-u$ and $\bar
z=i\,\ID_+\ID_-\bar u$.

\section{The Hopf surface: a canonical example}
Wess-Zumino-Witten models on even dimensional reductive Lie  group manifolds all allow for at least $N=(2,2)$ supersymmetry \cite{Spindel:1988nh}. A complex structure on the group is fully determined by its action on the Lie algebra and it is almost equivalent to a choice for a Cartan decomposition of the Lie algebra. Indeed on generators corresponding to positive (negative) roots it has eigenvalue $+i$ ($-i$) and the only freedom left is on its action on the Cartan subalgebra (CSA) which is mapped to itself and restricted by the condition that the invariant metric on the CSA is hermitian. 

The group $SU(2)\times U(1)$ is the canonical example.
As this model actually has an $N=(4,4)$ supersymmetry it has two 2-spheres of complex structures, one for each chirality. A detailed analysis \cite{Sevrin:2011mc} shows that this gives rise to two superspace formulations of the model: one in terms of a chiral and a twisted chiral superfield \cite{Rocek:1991vk} and one in terms of a semi-chiral multiplet \cite{Sevrin:1996jr}. 
 
We parameterize a group element $g\in SU(2)\times U(1)$,
\begin{eqnarray}
g= e^{i\rho }\,\left(
\begin{array}{cc} \cos \psi \, e^{i\varphi _1} & \sin \psi \,e^{i\varphi _2}\\
-\sin\psi\, e^{-i\varphi _2} & \cos\psi\, e^{-i\varphi _1}\end{array}\right),
\label{su2u1coor}
\end{eqnarray}
where $\varphi _1,\, \varphi_2,\, \rho \in \IR\,\mbox{mod}\,2 \pi $ and
$\psi\in[0, \pi /2]$. The group
manifold $S^3\times S^1$ can also be viewed as a rational Hopf surface defined by 
$\big(\IC^2\backslash(0,0)\big)/\Gamma$ where elements of $\Gamma$ act on 
$(w,z)\in\IC^2\backslash(0,0)$ as $(w,z) \rightarrow e^{2 \pi n}(w,z)$, $n\in\IZ$, which is clear from the following identification,
\begin{eqnarray}
w&=&\cos\psi\,e^{-\rho-i \varphi_1}, \nonumber\\
z&=&\sin\psi\,e^{-\rho+i \varphi_2}\,.\label{coorch}
\end{eqnarray}
It turns out that these coordinates can be identified with a twisted chiral superfield $w$ and a chiral superfield $z$. The resulting potential is given by,
\begin{eqnarray} V_{w\neq 0}(z,
w, \bar z, \bar w) = \int^{\frac{z \bar z}{ w \bar w}} \frac{dq}{q}
\ln\left(1+q\right) - \frac{1}{2} \left( \ln w \bar w \right)^2 \,
,\label{su2u1ctc4moduli}
\end{eqnarray}
which is well defined as long as $w\neq 0$ (or $\psi\neq \pi/2$). The 
``mirror'' potential (see eq.~(\ref{mirror})),
\begin{eqnarray} V_{z\neq 0}(w,
\bar w, z,\bar z) = -\int^{\frac{w \bar w}{ z \bar z}} \frac{dq}{q}
\ln\left(1+q\right) + \frac{1}{2} \left( \ln z \bar z \right)^2 \,,\label{WZWpot2b}
\end{eqnarray}
is well defined as long as $z\neq 0$ (or $\psi\neq 0$). On the overlap of the two patches the potentials are related by a generalized K\"ahler transformation,
\begin{eqnarray}
V_{w\neq 0}-V_{z\neq 0}=-\ln\big( z\bar z\big)\ln\big( w\bar w\big)\,.
\end{eqnarray}
The type is $(k_+,k_-)=(1,1)$ and no type changing occurs. The closed pure spinors can be calculated \cite{Sevrin:2011mc} and their Mukai pairings are given by,
\begin{eqnarray}
\big( \phi_+, \bar\phi_+\big)=\big( \phi_-, \bar \phi_-\big)=-\, \frac{2}{z \bar{z}
+w \bar{w}}\,,
\end{eqnarray}
which vanish nowhere and which  satisfy the generalized Calabi-Yau condition eq.~(\ref{CYcd1}). However a more careful investigation shows that they are not globally well defined defined implying that the current model does not provide an appropriate supergravity solution. 

The second formulation of the model in terms of a semi-chiral multiplet is given by,
\begin{eqnarray}
V_{\psi\neq 0} = \ln \frac{l}{ \bar r}\,\ln \frac{\bar l}{ r}\,-\int^{r\bar r}\frac{dq}{q}\,
 \ln\big(1+q\big)\,,
\end{eqnarray}
where the semi-chiral coordinates are related to the previous one by,
\begin{eqnarray}
 l=w\,,\qquad \bar l=\bar w\,,\qquad r=\frac{\bar w}{z}\,,\qquad \bar r=\frac{ w}{\bar z}\,.\label{simprel}
\end{eqnarray}
In order to obtain a potential which is well defined in $\psi =0$ we perform a Legendre transformation,
\begin{eqnarray}
V_{\psi\neq 0}(l,\bar l,r,\bar r)-\ln l\,\ln l'-\ln \bar l \, \ln \bar l'\,,
\end{eqnarray}
where $l$ is now unconstrained and $l'$ satisfies $\ID_+ l'=0$. Varying w.r.t. $l$ and $\bar l$ one finds that $l'=\bar l/r$ and $\bar l'= l/\bar r$. 
Making a final coordinate transformation $r'=1/r$ and $\bar r'=1/\bar r$ we get,
\begin{eqnarray}
V_{\psi\neq 0}(l,\bar l,r,\bar r)-\ln l\,\ln l'-\ln \bar l \, \ln \bar l'=
V_{\psi\neq \frac{\pi}{2}}(l',\bar l',r',\bar r')-\frac 1 2 \,\big(\ln r'\big)^2-\frac 1 2 \,\big(\ln \bar r'\big)^2\,,
\end{eqnarray}
where the last two terms can be removed by a generalized K\"ahler transformation. The resulting potential which is now well defined in $\psi=0$ is
explicitly given by,
\begin{eqnarray}
V_{\psi\neq \frac{\pi}{2}}=-\ln \frac{l'}{  r'}\,\ln \frac{\bar l'}{\bar r'}\,+\int^{r'\bar r'}\frac{dq}{q}\,
 \ln\big(1+q\big)\,,
\end{eqnarray}
and it is once more the mirror transform of the original potential. This is an explicit example of a situation where the generalized potentials on the overlap of two coordinate patches are not only related by a generalized K\"ahler transformation but by a Legendre transformation as well. 

Using the potentials to calculate the complex structures one verifies that type-changing occurs at $\psi=\pi/2$ where the type jumps as $(0,0) \rightarrow (0,2)$ and at $\psi=0$ where we have that the type goes as $(0,0) \rightarrow (2,0)$. The generalized Calabi-Yau condition eq.~(\ref{CYcd1}) does not hold anymore however one verifies that the weaker UV-finiteness condition eq.~(\ref{CYcd2})  is still satisfied.

The formulation of the model in terms of a pair of twisted and chiral superfields and the one in terms of a semi-chiral multiplet are related by a T-duality transformation (as originally suggested in \cite{martinWIP} and worked out explicitly in \cite{Sevrin:2011mc}). Indeed, performing a T-duality transformation along the $S^1$ of $S^1\times S^3$ brings one back to $S^1\times S^3$. The isometry which will be gauged acts as $\rho \rightarrow \rho+\mbox{constant}$ which is easily verified to be of the type discussed in section 5.1. This T-duality transformation provides an explanation for the simple relation between the chiral/twisted chiral and the semi-chiral parameterization eq.~(\ref{simprel}).  

\section{Discussion and outlook}
In the current review we limited ourselves to non-linear $\sigma$-models without boundaries. However boundaries can be added to  $N=(2,2)$ superspace which gives rise to an intricate and powerful worldsheet description of supersymmetric D-branes including configurations interpolating between A-branes and B-branes \cite{Sevrin:2008tp}, \cite{Sevrin:2009na}.

The previous shows that supersymmetric non-linear $\sigma$-models exhibit a rich geometric structure, both at the classical and at the quantum level. Mathematically the resulting bi-hermitian geometry is perfectly captured by Hitchin's generalized K\"ahler geometry. The analysis of the one loop quantum corrections leads to expressions closely related to Hitchin's generalized Calabi-Yau geometry. Knowing that higher loop corrections result in a deformation of the one-loop $\beta$-functions, a natural and interesting question arises which is whether those can be put in a generalized geometrical setting as well, leading to a better and more systematic control of the perturbation series. 

Another interesting question deals with the doubled formalism \cite{Hull:2004in}.  The doubled formalism provides a concrete proposal to tackle non-geometric or T-fold compactifications.  Although the underlying geometry is still being developed,  hints at a very rich framework beyond conventional Riemannian geometry are emerging (see for instance  \cite{Hohm:2012mf}). Many expressions are reminiscent of, and indeed motivated by,  Hitchin's generalized complex geometry. As supersymmetric non-linear $\sigma$-models in $N=(2,2)$ superspace provide a very explicit local realization of generalized K\"ahler geometry, one could expect that the development of a doubled formalism in $N=(2,2)$ superspace will shed light on these novel geometric structures. Such a doubled formalism certainly exists but has a slightly different structure from what is customary.

A simple example of a $\sigma$-model on $T^2$ illustrates this point. The potentials $V=(z+\bar z)^2/2$ and $W=-(w+\bar w)^2/2$ are T-dual. In a manifest $N=(2,2)$ supersymmetric doubled formalism one has to "overdouble" the coordinates, {\em i.e.} all superfields involved in the T-duality transformation are doubled. In this case the doubled formalism is parameterized by $z$, $\bar z$, $w$ and $\bar w$. The overdoubled coordinate is eliminated by the condition,
\begin{eqnarray}
w+\bar w = z+\bar z\,,\label{coiso}
\end{eqnarray}
which singles out a kind of co-isotropic brane in the overdoubled space. Hull's constraints follow immediately from eq.~(\ref{coiso}),  indeed,
\begin{eqnarray}
\hat D_\pm(w+\bar w)= \hat D_\pm (z+\bar z) \Rightarrow D_\pm (w-\bar w)=\pm
D_\pm(z-\bar z)\,.
\end{eqnarray}
The expression above implies that we are dealing with superfields which are truly chiral: from eq.~(\ref{coiso}) we get that $\ID_+(z-w)=\ID_-(z-\bar w)=0$. From the normal superfield constraints we get   $\bar\ID_+(z-w)=\bar\ID_-(z-\bar w)=0$ implying that $\partial _\pp(z-w)=\partial _=(z-\bar w)=0$
as well. Chiral bosons in the sense just mentioned are currently being developed together with a  full treatment of the doubled formalism in $N=(2,2)$ superspace \cite{wip}.

Finally, it would be most useful to have some more non-trivial examples of generalized K\"ahler geometries where the local structure -- read the generalized K\"ahler potential -- is explicitly known. WZW-models on even dimensional reductive Lie group manifolds might provide them however at this point explicit results have only been obtained for the $SU(2)\times U(1)$ and the $SU(2)\times SU(2)$ models.

\section*{Acknowledgments}

AS would like to thank the organizers of the Corfu Summer Institute and the XVIIIth European Workshop on String Theory for the nice scientific environment. This work is supported by the FWO-Vlaanderen, Project G.0114.10N, by the Belgian Federal Science Policy Office through the Interuniversity Attraction Pole P7/37 and by the Vrije Universiteit Brussel through the Strategic Research Program ``High-Energy Physics''. DT is a Postdoctoral Researcher FWO-Vlaanderen.

\appendix
\section{Notation and conventions}
 We denote the worldsheet coordinates by $ \tau,\sigma \in\IR$, and the worldsheet light-cone coordinates are
defined by,
\begin{eqnarray}
\sigma ^\pp= \tau + \sigma ,\qquad \sigma ^== \tau - \sigma .\label{App1}
\end{eqnarray}
The $N=(1,1)$ (real) fermionic coordinates are denoted by $ \theta ^+$ and $ \theta ^-$ and the
corresponding derivatives satisfy,
\begin{eqnarray}
D_+^2= - \frac{i}{2}\, \partial _\pp \,,\qquad D_-^2=- \frac{i}{2}\, \partial _= \,,
\qquad \{D_+,D_-\}=0.\label{App2}
\end{eqnarray}
The $N=(1,1)$ integration measure is explicitely given by,
\begin{eqnarray}
\int d^ 2 \sigma \,d^2 \theta =\int d\tau \,d \sigma \,D_+D_-.
\end{eqnarray}
Passing from $N=(1,1)$ to $ N=(2,2)$ superspace requires
the introduction of two more real fermionic coordinates $ \hat \theta ^+$ and $ \hat \theta ^-$
where the corresponding fermionic derivatives satisfy,
\begin{eqnarray}
\hat D_+^2= - \frac{i}{2} \,\partial _\pp \,,\qquad \hat D_-^2=- \frac{i}{2} \,\partial _= \,,\label{app4}
\end{eqnarray}
and again all other -- except for (\ref{App2}) -- (anti-)commutators do vanish.
The $N=(2,2)$ integration measure is,
\begin{eqnarray}
\int d^2 \sigma \,d^2 \theta \, d^2 \hat \theta =
\int d \tau\, d \sigma \,D_+D_-\, \hat D_+ \hat D_-.
\end{eqnarray}
Regularly a complex basis is used,
\begin{eqnarray}
\ID_\pm\equiv \hat D_\pm+i\, D_\pm,\qquad
\bar \ID_\pm\equiv\hat D_\pm-i\,D_\pm,
\end{eqnarray}
which satisfy,
\begin{eqnarray}
\{\ID_+,\bar \ID_+\}= -2i\, \partial _\pp\,,\qquad
\{\ID_-,\bar \ID_-\}= -2i\, \partial _=,
\end{eqnarray}
and all other anti-commutators do vanish.

\end{document}